\documentclass[aps,twocolumn,twoside,nofootinbib,superscriptaddress,
               preprintnumbers]{revtex4}

\usepackage{amsmath,amsthm}

\renewcommand{\>}{\rangle}

\newcommand{\be}{\begin{equation}}
\newcommand{\ee}{\end{equation}}
\newcommand{\nn}{\nonumber\\}

\renewcommand{\P}{{\cal P}}
\renewcommand{\S}{{\cal S}}

\newcommand{\nint}[1]{\lfloor #1 \rceil}

\newtheorem{theorem}{Theorem}
\newtheorem{lemma}[theorem]{Lemma}
\newtheorem*{problem}{Problem}
\newtheorem*{conjecture}{Conjecture}

\newdimen\arrayruleHwidth
\makeatletter
\def\Hline{\noalign{\ifnum0=`}\fi\hrule \@height \arrayruleHwidth
  \futurelet \@tempa\@xhline}
\makeatother

\begin{document}

\title{Quantum algorithms for subset finding}

\author{Andrew M. Childs}
\email[]{amchilds@mit.edu}

\affiliation{Center for Theoretical Physics,
             Massachusetts Institute of Technology,
             Cambridge, MA 02139, USA}

\author{Jason M. Eisenberg}
\email[]{jme@mit.edu}

\affiliation{Center for Theoretical Physics,
             Massachusetts Institute of Technology,
             Cambridge, MA 02139, USA}

\affiliation{Computer Science and Artificial Intelligence Laboratory,
             Massachusetts Institute of Technology,
             Cambridge, MA 02139, USA}

\date[]{1 December 2003}

\preprint{MIT-CTP \#3438}


\begin{abstract}
Recently, Ambainis gave an $O(N^{2/3})$-query quantum walk algorithm
for element distinctness, and more generally, an
$O(N^{L/(L+1)})$-query algorithm for finding $L$ equal numbers.  We
point out that this algorithm actually solves a much more general
problem, the problem of finding a subset of size $L$ that satisfies
any given property.  We review the algorithm and give a considerably
simplified analysis of its query complexity.  We present several
applications, including two algorithms for the problem of finding an
$L$-clique in an $N$-vertex graph.  One of these algorithms uses
$O(N^{2L/(L+1)})$ edge queries, and the other uses $\tilde
O(N^{(5L-2)/(2L+4)})$, which is an improvement for $L \le 5$.  The
latter algorithm generalizes a recent result of Magniez, Santha, and
Szegedy, who considered the case $L=3$ (finding a triangle).  We also
pose two open problems regarding continuous time quantum walk and
lower bounds.
\end{abstract}

\maketitle

\section{Introduction}

In this paper, we address the following rather general query
complexity problem:
\begin{problem}[$\boldsymbol{L}$-subset finding] ~
\begin{description}
\item Input:
{\em ($i$)} Black box function $f:D \to R$, where the domain $D$ and the
range $R$ are finite sets, and $|D|=N$ is the problem size.
{\em ($ii$)} Property $\P \subset (D \times R)^L$.

\item Output: 
Some $L$-subset $\{x_1,\ldots,x_L\} \subset D$ such that
$((x_1,f(x_1)),\ldots,(x_L,f(x_L)))\in\P$, or {\sc reject} if none
exists.
\end{description}
\end{problem}
\noindent
In the general case, for the purpose of query complexity, we can
assume that the property $\P$ is given explicitly as a list.  However,
for specific versions of this problem, there will generally be a much
more compact way to specify the property.

The quantum query complexity of the $L$-subset finding problem is
well-understood in the cases $L=1,2$.  The 1-subset finding problem is
nothing but the well-known unstructured search problem.  This problem
can be solved in $O(\sqrt N)$ queries~\cite{Gro97}, which is
optimal~\cite{BBBV97}.  The element distinctness problem
(see~\cite{Buh00} and references therein) is a particular case of
2-subset finding.  This problem can be solved using $O(N^{2/3})$
queries using a recent algorithm of Ambainis~\cite{Amb03b}, which is
also optimal~\cite{Aar02,Shi01,Amb03a,Kut03}.

In~\cite{Amb03b}, Ambainis also gave an $O(N^{L/(L+1)})$-query
algorithm for the $L$-element distinctness problem, where the goal is
to find $L$ inputs that give the same output.  This is a particular
case of $L$-subset finding with $\P=\{((x_1,y),\ldots,(x_L,y))\}$.
However, the algorithm is actually even more general: it solves the
$L$-subset finding problem using $O(N^{L/(L+1)})$ queries, regardless
of $\P$.  We review the algorithm in this context in Section
\ref{sec:algorithm}.  We also present a simplified proof that the
algorithm works in Section \ref{sec:analysis}.

Note that for $L$-element distinctness, any $L$-subset of
inputs can potentially satisfy $\P$, so we could have described $\P$
as a subset of $R^L$.  Many variants of element distinctness could
also be described this way.  However, for other problems such as
finding a clique in a graph, we require the generality provided by
letting $\P \subset (D \times R)^L$.

Because the property $\P$ can be any subset of $(D \times R)^L$, the
subset finding algorithm can be applied to a wide variety of related
problems.  We discuss some of these applications in Section
\ref{sec:applications}.  In particular, we consider using the
algorithm to find $L$-vertex subgraphs in $N$-vertex graphs, assuming
the graph is given as a black box allowing edge queries.  The most
straightforward application of the $L$-subset finding algorithm gives
an $O(N^{2L/(L+1)})$-query algorithm for finding an $L$-clique.  This
upper bound is the best we know for $L \ge 6$.  However, for $L \le 5$
we improve it to $\tilde O(N^{(5L-2)/(2L+4)})$ using the recursive
approach of~\cite{MSS03}, which considers the particular case $L=3$
(finding a triangle).

We conclude in Section \ref{sec:problems} by suggesting some open
problems.  We discuss why the subset finding algorithm is well-suited
to discrete rather than continuous time quantum walks, and we pose an
open problem related to the simulation of continuous time quantum
walks.  We also discuss the known lower bounds for $L$-subset finding,
and we suggest a specific case where the subset finding algorithm
might be optimal for general $L$.

\section{Algorithm}
\label{sec:algorithm}

In this section, we review Ambainis's algorithm for element
distinctness~\cite{Amb03b} in the context of subset finding.  The
algorithm is based on the idea of a discrete time quantum walk on a
graph~\cite{Mey96,Wat01,AAKV01,ABNVW01}.  A quantum walk is simply a
way of formulating local quantum dynamics on a graph.  If the walk
takes discrete steps that only move amplitude between neighboring
vertices, then in general the walk must include an ancillary state
space~\cite{Mey96}, sometimes referred to as a ``coin,'' that can be
used to indicate the direction of the next step of the walk.  Thus the
quantum walk can be built as a sequence of transformations, some of
which act on the coin register and some of which use the state of the
coin register to update the location in the graph.

The graph used in Ambainis's construction is a bipartite graph whose
vertices are all subsets of the domain $D$ of size either $M$ or
$M+1$.  We will choose $M$ to be $\Theta(N^q)$ for some $0 < q < 1$
(recall $|D|=N$).  Let $A \subset D$ with $|A|=M$, and let $B \subset
D$ with $|B|=M+1$.  Vertices $A$ and $B$ are connected in $G$ iff $|A
\cap B| = M$, i.e., $B = A \cup \{k\}$ for some $k \in D \setminus A$.

To define the quantum walk on $G$, we define an orthonormal basis of
quantum states $|A\>$, one for each subset $A$.  For any subset $A$
with $|A|=M$, there are $M$ associated function values $f(A) \in R^M$,
and similarly, for any subset $B$ with $|B|=M+1$, there are $M+1$
associated function values $f(B) \in R^{M+1}$.  A key idea of the
algorithm is to store these function values along with the subset.
The only parts of the algorithm that require queries will be those
that manipulate the function values.

The full state of the quantum computer has the form $|A,f(A),k\>$
where $|A,f(A)\>$ is the state described previously, including the
function values, and $|k\>$ is the coin register, where $k \in D$.  If
$|A|=M$ then $k$ indicates an element to be added to $A$, so we must
have $k \notin A$.  Similarly, if $|B|=M+1$ then $k$ indicates an
element to be removed from $B$, so we must have $k \in B$.

One step $W$ of the quantum walk (which actually involves two steps on
$G$) is a product of four unitary transformations, $W=S C_2 S C_1$.
The shift operation $S$ acts as
\begin{align}
  S |A,f(A),k\> &= |A \cup \{k\},f(A \cup \{k\}),k\> \\
  S |B,f(B),k\> &= |B \setminus \{k\},f(B \setminus \{k\}),k\>
\end{align}
and can be implemented using one query of the black box.  The coin
operations $C_1$ and $C_2$ are Grover diffusion operators on $k \notin
A$ and $k \in B$, respectively.  In other words, we have
\begin{align}
  C_1 |A,f(A),k\> &= |A,f(A),k\> \nn
    & \quad - \frac{2}{N-M} \sum_{k' \notin A} |A,f(A),k'\> \\
  C_1 |B,f(B),k\> &= |B,f(B),k\>
\end{align}
and
\begin{align}
  C_2 |A,f(A),k\> &= |A,f(A),k\> \\
  C_2 |B,f(B),k\> &= |B,f(B),k\> \nn
    & \quad - \frac{2}{M+1} \sum_{k' \in B} |B,f(B),k'\>
\,.
\end{align}
These unitary transformations do not change the subset, so they also do
not affect the function values, and consequently do not require any
queries.

The algorithm also involves a phase flip operation that distinguishes
subsets $A$ that include an $L$-subset satisfying $\P$.  For
simplicity, we suppose there is exactly one special subset
$\S=\{x_1,\ldots,x_L\} \subset D$ of inputs such that
$((x_1,f(x_1)),\ldots,(x_L,f(x_L)))\in\P$.  The general case can be
handled by modifying the algorithm using standard sampling
techniques~\cite{Amb03b}.  The phase flip operation is
\be
  P |A,f(A)\> =
  \begin{cases}
    -|A,f(A)\> & \S \subset A \\
     |A,f(A)\> & \S \not\subset A \,.
  \end{cases}
\label{eq:phaseflip} 
\ee
Given $A$ and $f(A)$, the property $\P$ can be checked without any
further queries of the black box, so no queries are required to
implement $P$.

The initial state for the algorithm is the symmetric state on subsets
of size $M$,
\be
  |s\> =
  \frac{1}{\sqrt{c}}
    \sum_{|A|=M} |A,f(A)\>
    \sum_{k \notin A} |k\> 
\,,
\ee
where $c=\binom{N}{M} (N-M)$ is a normalization constant.  This state
can be made using $M$ queries to the black box.

The full algorithm is $(W^{t_1} P)^{t_2}$, which uses $2 t_1 t_2$
additional queries, where $t_1,t_2$ will be determined in the
analysis.  Thus, using $M+2t_1t_2$ queries, we produce the state
$(W^{t_1} P)^{t_2} |s\>$.  Our goal is to choose $t_1,t_2$ so that a
measurement of this state is likely to provide a solution to the
problem.

Note that in the description of this algorithm, we have focused only
on the query complexity of the various steps.  For the algorithm to be
efficient in terms of the number of computational steps as well as the
number of queries, it must be possible to use $A$ and $f(A)$ to
efficiently determine whether there is an $\{x_1,\ldots,x_L\} \subset
A$ such that $((x_1,f(x_1)),\ldots,(x_L,f(x_L)))\in\P$.  In doing so
it may be helpful to maintain the function values in some particular
data structure, such as a hash table for the element distinctness
problem (see Section 6 of~\cite{Amb03b}).  However, for the general
problem we will not be concerned with how $\P$ is given or how
efficiently it can be checked.

\section{Analysis of the algorithm}
\label{sec:analysis}

In this section, we give an analysis of the subset finding algorithm
that is simpler, as well as tighter, than that of~\cite{Amb03b}.  The
main result is the following:
\begin{theorem}
  The quantum query complexity of $L$-subset finding is $O(N^{L/(L+1)})$.
\label{thm:mainresult}
\end{theorem}

To prove this theorem, we need to understand the spectrum of $W^{t_1}
P$, which we will see is close to a small rotation in a
two-dimensional subspace spanned by $|s\>$ and a state that gives a
solution to the problem.  We will begin by proving
Lemma~\ref{lemma:walkspect} below, which characterizes the spectrum of
the walk step $W$ alone and shows how to choose $t_1$.  Then we will
use techniques from~\cite{CG03} to prove Lemma~\ref{lemma:algspect}
below, which describes the spectrum of $W^{t_1} P$ and shows how to
choose $t_2$.  Finally, we will give the proof of Theorem
\ref{thm:mainresult}.

To analyze the algorithm, we will use the fact that the evolution
takes place in a $(2L+1)$-dimensional subspace of the full Hilbert
space.  Let $\S_1=\S$ and $\S_0=D \setminus \S$.  Define the states
\be
  |A_{j,p}\> = \frac{1}{\sqrt{c_{j,p}}} 
                 \sum_{\substack{|A|=M \\ |A \cap \S|=j}} |A,f(A)\>
                 \sum_{\substack{k \notin A \\ k \in \S_p}} |k\>
\ee
for $j=0,1,\ldots,L-1$, $p=0,1$ and for $j=L$, $p=0$, and the states
\be
  |B_{j,p}\> = \frac{1}{\sqrt{d_{j,p}}} 
                 \sum_{\substack{|B|=M+1 \\ |B \cap \S|=j}} |B,f(B)\>
                 \sum_{\substack{k \in B \\ k \in \S_p}} |k\>
\ee
for $j=0$, $p=0$ and for $j=1,\ldots,L$, $p=0,1$.  Here the
normalization constants are given by
\begin{align}
  c_{j,0} &= \binom{N-L}{M-j}   \binom{L}{j} [(N-L)-(M-j)] \\
  c_{j,1} &= \binom{N-L}{M-j}   \binom{L}{j} (L-j) \\
  d_{j,0} &= \binom{N-L}{M+1-j} \binom{L}{j} (M+1-j) = c_{j,0} \\
  d_{j,1} &= \binom{N-L}{M+1-j} \binom{L}{j} j       = c_{j-1,1}
\,.
\end{align}

In terms of these states, the shift operation acts as
\begin{align}
  S|A_{j,0}\> &= |B_{j,0}\> \\
  S|A_{j,1}\> &= |B_{j+1,1}\> \\
  S|B_{j,0}\> &= |A_{j,0}\> \\
  S|B_{j,1}\> &= |A_{j-1,1}\>
\,.
\end{align}
By explicit calculation, the coin transformations have the following
matrix elements (with others determined by the fact that $C_1,C_2$ are
Hermitian as well as unitary):
\begin{align}
  \<A_{j,0}|C_1|A_{j,0}\> &= 
                             1-2\alpha(L-j) \\
  \<A_{j,0}|C_1|A_{j,1}\> &= 
                             2\sqrt{\alpha(L-j)[1-\alpha(L-j)]} \\
  \<A_{j,1}|C_1|A_{j,1}\> &= 
                             2\alpha(L-j)-1
\end{align}
where $\alpha=1/(N-M)$, and
\begin{align}
  \<B_{j,0}|C_2|B_{j,0}\> &= 
                             1-2\beta j \\
  \<B_{j,0}|C_2|B_{j,1}\> &= 
                             2\sqrt{\beta j(1-\beta j)} \\
  \<B_{j,1}|C_2|B_{j,1}\> &= 
                             2\beta j-1
\end{align}
where $\beta=1/(M+1)$.

Now we are ready to analyze the spectrum of the walk step $W=S C_2 S
C_1$.  Similarly to Lemma~2 of~\cite{Amb03b}, we find
\begin{lemma}
  The eigenvalues of $W$ are $1$ and $\exp[\pm i (2\sqrt{j/M} +
  O(1/M+1/\sqrt N))]$ for $j=1,\ldots,L$.  The eigenvector with
  eigenvalue $1$ is $|s\>$, and the two eigenvectors with eigenvalues
  $\exp[\pm i (2\sqrt{L/M} + O(1/M+1/\sqrt N))]$ are
  $\frac{1}{\sqrt 2}(|A_{L-1,1}\> \pm i |A_{L,0}\>)+O(\sqrt{M/N})$.
\label{lemma:walkspect}
\end{lemma}
\noindent
In fact, the eigenvalues of $|W-1|$ are exactly $0$ and
$\sqrt{j(\alpha+\beta-j\alpha\beta)}$ for $j=1,2,\ldots,L$, but we
will not require this level of detail.

Note that the notation $|\psi\> = |\phi\> + O(\epsilon)$ means $\|
|\psi\>-|\phi\> \| = O(\epsilon)$.  Similarly, for operators $X,Y$ we
write $X=Y+O(\epsilon)$ as shorthand for $\| X-Y \| = O(\epsilon)$.

\begin{proof}
Direct calculation shows that $|s\>$ is an eigenvector of $W$ with
eigenvalue $1$.

To understand the rest of the spectrum, we can think of the walk as
being composed of two parts, $C_1$ and $S C_2 S$, each of which is a
unitary transformation in the $A$ subspace.  Let
\begin{align}
  C_1     &= C + \Delta_1 \\
  S C_2 S &= C + \Delta_2 
\end{align}
where $C$ is the diagonal matrix with diagonal elements
\be
  \<A_{j,p}|C|A_{j,p}\> = (-1)^p
\,.
\ee
The matrices $\Delta_1,\Delta_2$ each consist of $L$ $2\times2$ blocks
and one $1\times1$ block (a zero), so their eigenvalues can be
computed explicitly, giving $\|\Delta_1\| = O(1/\sqrt N)$ and
$\|\Delta_2\| = O(1/\sqrt M)$.  Now we have
\begin{align}
  W &= 1 + \Delta_2 C + C \Delta_1 + \Delta_2 \Delta_1 \\
    &= 1 + \Delta_2 C + O(1/\sqrt{N})
\,.
\end{align}
Therefore, it is sufficient to calculate the eigenvectors and
eigenvalues of the matrix $\Delta_2 C$, which has matrix elements
\begin{align}
  \<A_{j,0}  |\Delta_2 C|A_{j,0}\>   &= -2\beta j \\
  \<A_{j,0}  |\Delta_2 C|A_{j-1,1}\> &= -2\sqrt{\beta j(1-\beta j)} \\
  \<A_{j-1,1}|\Delta_2 C|A_{j,0}\>   &= 2\sqrt{\beta j(1-\beta j)} \\
  \<A_{j-1,1}|\Delta_2 C|A_{j-1,1}\> &= -2\beta j
\,.
\end{align}
This matrix also has $L$ $2 \times 2$ blocks and one $1 \times 1$
block, so its eigenvalues and eigenvectors can also be computed
explicitly.  Using perturbation theory to calculate the size of the
small correction from the $O(1/\sqrt N)$ term, we find that the
eigenvalues of $W-1$ are $0$ and $\pm2i\sqrt{j\beta} + O(1/\sqrt N)$
for $j=1,\ldots,L$, which implies the claim about the eigenvalues of
$W$.  The claim about the two relevant eigenvectors of $W$ also
follows by perturbation theory.
\end{proof}

Lemma~\ref{lemma:walkspect} shows that in $W^{t_1}$ we want to take
$t_1 = O(\sqrt M)$, since this choice implements a rotation by an
angle of order unity.  In fact,  we choose $t_1 = \nint{\frac{\pi}{2}
\sqrt{M/L}}$, where $\nint x$ denotes the nearest integer to $x$, so
that the two extreme eigenvalues of $W^{t_1}$ are $-1 + O(1/\sqrt{M})$
(here the dominant source of error is actually the fact that
$\frac{\pi}{2} \sqrt{M/L}$ may not be close to an integer).

We now describe techniques for understanding the spectrum of $W^{t_1}
P$.  Essentially the same analysis appears in Section IV.A
of~\cite{CG03}, and we will follow that treatment closely.  Our goal
is to calculate spectral properties of an operator $UP$ where
$P=1-2|w\>\<w|$.  In our case $U$ is $W^{t_1}$, and $P$ is given by
(\ref{eq:phaseflip}), so that $|w\>=|A_{L,0}\>$.

An eigenvector $|\theta_a\>$ of $UP$, with eigenvalue $e^{i\theta_a}$,
satisfies
\be
  UP |\theta_a\> = (U-2U|w\>\<w|) |\theta_a\> 
               = e^{i\theta_a}|\theta_a\>
\,,
\ee
i.e.,
\be
  (U-e^{i\theta_a}) |\theta_a\> = 2 U|w\>\<w|\theta_a\>
\,.
\ee
Define
\be
  R_a = |\<w|\theta_a\>|^2
\ee
and choose the phase of $|\theta_a\>$ so that
\be
  \<w|\theta_a\> = \sqrt{R_a}
\,.
\label{eq:ra}
\ee
As long as $R_a > 0$, we may write
\be
  |\theta_a\> = \frac{2 \sqrt{R_a}}{1 - U^\dag e^{i\theta_a}} |w\>
\,.
\label{eq:evec}
\ee
Consistency with (\ref{eq:ra}) then gives the eigenvalue condition
\be
  \<w| \frac{2}{1-U^\dag e^{i\theta_a}} |w\> = 1
\,.
\label{eq:evalcond}
\ee
Now let the eigenstates of $U$ be $|u_j\>$ with eigenvalues $e^{i
u_j}$.  Then the eigenvalue condition (\ref{eq:evalcond}) may be
written
\be
  \sum_j \frac{2|\<w|u_j\>|^2}{1-e^{i(\theta_a-u_j)}} = 1
\,.
\ee
The real part of this equation is automatically satisfied, and the
imaginary part gives
\be
  \sum_j |\<w|u_j\>|^2 \, \cot \left(\frac{\theta_a-u_j}{2}\right) 
  = 0
\,.
\ee
Roots of this equation determine the values of $\theta_a$, which
specify the eigenvalues of $UP$.

Similar considerations let us determine properties of the
eigenvectors.  The normalization condition on $|\theta_a\>$ gives
\be
  R_a \<w| \frac{4}{|1-U^\dag e^{i\theta_a}|^2} |w\> = 1
\,,
\ee
i.e.,
\begin{align}
  R_a &= \left[ \sum_j 
         \frac{4 |\<w|u_j\>|^2}{|1-e^{i(\theta_a-u_j)}|^2}
         \right]^{-1} \\
      &= \left[1+\sum_j |\<w|u_j\>|^2 \, 
         \cot^2 \left(\frac{\theta_a-u_j}{2}\right)
         \right]^{-1}
\,.
\label{eq:woverlap}
\end{align}
To compute the overlap of $|\theta_a\>$ on the eigenstates $|u_j\>$ of
$U$, we multiply (\ref{eq:evec}) on the left by $\<u_j|$, which gives
\be
  \<u_j|\theta_a\> 
  = \<w|\theta_a\>\<u_j|w\>
    \left[1+i \cot\left(\frac{\theta_a-u_j}{2}\right)\right]
\label{eq:eigoverlap}
\,.
\ee

Applying these techniques to our specific problem, we find
\begin{lemma}
  $W^{t_1} P$ has two eigenvectors $|\theta_\pm\> = \frac{1}{\sqrt 2}
  (|w\> \pm i |s\>) + O(1/M+M/N)$ with eigenvalues $\exp[\pm 2i
  (M/N)^{L/2} (1+O(1/M+M/N))]$.
\label{lemma:algspect}
\end{lemma}

\begin{proof}
From (\ref{eq:phaseflip}), we have $|w\> = |A_{L,0}\>$.  The
eigenvalue condition (\ref{eq:evalcond}) for $W^{t_1}P$ is
\be
  0 = |\<w|s\>|^2 \, \cot\frac{\theta_a}{2} 
        + \sum_{j \ne 0} |\<w|u_j\>|^2 \, \cot\frac{\theta_a-u_j}{2}
\,.
\ee
We are interested in solutions where $\theta_a$ is small (which we
will indeed find for large $N$), so Taylor expansion in $\theta_a$
gives
\begin{align}
  0 &= |\<w|s\>|^2 \left(\frac{2}{\theta_a} + O(1)\right) \nn
    &\quad - \sum_{j \ne 0} |\<w|u_j\>|^2 \left(\cot\frac{u_j}{2} +
        \frac{\theta_a}{2} \csc^2\frac{u_j}{2} + O(\theta_a^2)\right) \\
    &= |\<w|s\>|^2 \left(\frac{2}{\theta_a} + O(1)\right) \nn
    &\quad - \sum_{\text{pairs of } j} |\<w|u_j\>|^2
        \left(\theta_a\csc^2\frac{u_j}{2} + O(\theta_a^2)\right)
\end{align}
where in the second equality we have used the fact that the values
$u_j$ come in $\pm$ pairs, with identical values of $|\<w|u_j\>|^2$,
since $W$ is a real unitary matrix.  From Lemma~\ref{lemma:walkspect},
we see that the only nonnegligible contributions to this condition
come from $|s\>$ and the two eigenstates with $|\<w|u_j\>|^2 =
\frac{1}{2} + O(M/N)$.  We have
\begin{align}
  0 &= |\<w|s\>|^2 \left(\frac{2}{\theta_a}+O(1)\right) \nn
    &\quad -\left(\frac{1}{2} + O(M/N)\right) \theta_a [1+O(1/M)]
           + O(\theta_a^2) \nn
    &\quad + O(M/N) \, O(\theta_a)
\,,
\end{align}
which gives two solutions,
\be
  \theta_{\pm} = \pm 2\<w|s\> + O((M/N)^{1+L/2})
\,,
\ee
where the leading order contribution as well as the error term follow from
\begin{align}
  \<w|s\> &= \sqrt{c_{L,0}/c} \\
          &= \sqrt\frac{(N-L)! M!}{N! (M-L)!} \\
          &= (M/N)^{L/2} \, [1+O(1/M)]
\,.
\end{align}

Having found the eigenvalues of $W^{t_1} P$, we can now calculate its
eigenstates.  Applying Lemma~\ref{lemma:walkspect} to
(\ref{eq:woverlap}) gives
\be
  \<w|\theta_\pm\> = \frac{1}{\sqrt 2}+O(1/M+M/N)
\,,
\ee
and a similar calculation using (\ref{eq:eigoverlap}) gives
\be
  \<s|\theta_\pm\> = \pm\frac{i}{\sqrt 2}+O(1/M+M/N)
\,,
\ee
which completes the proof.
\end{proof}

From Lemma~\ref{lemma:algspect}, we see that we should choose $t_2 =
\nint{\frac{\pi}{2} (N/M)^{L/2}}$ so that $(W^{t_1} P)^{t_2}$
implements a rotation from $|s\>$ to a state near $|w\>$.

We now complete the proof of the main result:
\begin{proof}[Proof of Theorem~\ref{thm:mainresult}]
Recall that $t_1 = \nint{\frac{\pi}{2} \sqrt{M/L}}$ and $t_2 =
\nint{\frac{\pi}{2} (N/M)^{L/2}}$.  From Lemma~\ref{lemma:algspect},
it is easy to see that
\be
  |\<w|(W^{t_1} P)^{t_2}|s\>|^2 = 1 - O(1/M+M/N)
\,.
\label{eq:finaloverlap}
\ee
Therefore, a measurement of the state $(W^{t_1} P)^{t_2}|s\>$ yields a
subset $A$ for which $\S \subset A$, together with the associated
function values $f(A)$, with probability close to $1$.  As discussed
in Section \ref{sec:algorithm}, the state $|s\>$ can be prepared using
$M$ queries, and the rest of the algorithm uses $2t_1t_2$ queries.
Thus the total query complexity is $M+2t_1 t_2=O(M+N^{L/2}
M^{(1-L)/2})$.  Choosing $M=\nint{N^{L/(L+1)}}$ shows that the
algorithm uses $O(N^{L/(L+1)})$ queries.
\end{proof}

Equation (\ref{eq:finaloverlap}) can be contrasted with Lemma~3
of~\cite{Amb03b}, which only shows that the success probability is
$O(1)$.

\section{Applications}
\label{sec:applications}

As discussed in~\cite{Amb03b}, the quantum walk subset finding
algorithm solves a natural generalization of element distinctness, the
problem of finding $L$ inputs to $f$ that give the same output, using
$O(N^{L/(L+1)})$ queries.  However, we have seen that the same
algorithm finds $L$-subsets satisfying an arbitrary property $\P$.
Therefore, we can also solve many variants of $L$-element distinctness
in $O(N^{L/(L+1)})$ queries---for example, finding a set of $L$
consecutive function values, relatively prime function values, or
sum-free function values (no two of the $L$ values sum to one of the
other $L$ values).

A closely related problem is that of finding an $L$-clique in an
$N$-vertex graph given an edge query oracle.  For $L=2$ this is simply
the problem of finding any edge in the graph, which is just the
unstructured search problem on $\binom{N}{2}$ items, and can be solved
using $\Theta(N)$ quantum queries.  For $L=3$ it is the problem of
finding a triangle, for which the exact quantum complexity is not
known.  A straightforward Grover search on the $\binom{N}{3}$ triples
of vertices gives an algorithm that uses $O(N^{3/2})$
queries~\cite{Buh00}, which was recently improved to $O(N^{10/7}
\log^2 N)$ in~\cite{Sze03} and subsequently to $\tilde O(N^{13/10})$
in~\cite{MSS03}.  However, the best known lower bound is only
$\Omega(N)$, which follows by straightforward reduction from the
unstructured search problem.  For $L>3$, the query complexity of
$L$-clique finding seems not to have been widely studied.

We describe two algorithms for finding $L$-cliques.  One algorithm
applies subset finding in a straightforward way, and another uses it
recursively.  The simple algorithm is faster than the recursive
algorithm for large $L$ ($L \ge 6$), and the recursive algorithm is
faster than the simple algorithm for small $L$ ($L \le 5$).  The
simple algorithm is a straightforward application of the quantum walk
subset finding algorithm where each subset $A$ over which we walk
consists of $M$ vertices, and where the algorithm also stores all of
the edges between them, i.e., their induced subgraph.  The analysis of
the algorithm proceeds as before, except now initialization requires
$O(M^2)$ queries to determine the subgraph, and each step of the walk
requires $O(M)$ queries to compute (or uncompute) the edges incident
on the newly added (or removed) vertex.  The total number of queries
is therefore
\be
  O(M^2+(N/M)^{L/2} \times \sqrt M \times M)
\,,
\label{eq:simple}
\ee
and choosing $M=\nint{N^{L/(L+1)}}$ gives an overall query complexity
of $O(N^{2L/(L+1)})$.  This result is no better than straightforward
Grover search for $L=3$, but it gives results that are better than
previously known algorithms for all fixed $L \ge 4$.

\begin{table}
\renewcommand{\baselinestretch}{1.5} \normalsize
\begin{tabular}{c@{~~~~~}r@{=}l@{~~~~~}r@{=}l}
  $L$ & \multicolumn{2}{c@{~~~~~}}{$\frac{2L}{L+1}$} 
      & \multicolumn{2}{c}{$\frac{5L-2}{2L+4}$} \\ \Hline
  2   & $\frac{4}{3}$  & 1.333   &   \multicolumn{2}{l}{1}   \\
  3   & $\frac{3}{2}$  & 1.5     &   $\frac{13}{10}$ & 1.3   \\
  4   & $\frac{8}{5}$  & 1.6     &   $\frac{3}{2}$   & 1.5   \\
  5   & $\frac{5}{3}$  & 1.667   &   $\frac{23}{14}$ & 1.643 \\ \hline
  6   & $\frac{12}{7}$ & 1.714   &   $\frac{7}{4}$   & 1.75  \\
  7   & $\frac{7}{4}$  & 1.75    &   $\frac{33}{18}$ & 1.833 \\
 $\vdots$ & \multicolumn{4}{c}{}
\end{tabular}
\renewcommand{\baselinestretch}{1} \normalsize
\caption{Query complexities of two algorithms for finding an
$L$-clique in an $N$-vertex graph.}
\label{tab:cliques}
\end{table}

However, we can improve upon this algorithm for $L \le 5$ using the
recursive approach from~\cite{MSS03}.  Here we again walk over induced
subgraphs of $M$ vertices, but now we search for a subgraph that
includes $L-1$ vertices from an $L$-clique in the full graph.  Recall
that to implement the phase flip $P$ we must determine whether a
subgraph satisfies this property.  If the subgraph does satisfy the
property, then either the $L$-clique lies entirely within the
subgraph, or all but one of the $L$-clique vertices form an
$(L-1)$-clique in the subgraph.  If the clique falls entirely within
the subgraph of size $M$, then no queries are necessary.  Otherwise,
if one of the vertices of the clique falls outside the subgraph, then
the clique can be found using a Grover search for the outside vertex,
where each Grover iteration uses $(L-1)$-subset finding to identify
the $L-1$ vertices inside the subgraph of size $M$.  The
implementation of the phase flip $P$ using this Grover search requires
a total of $r=\tilde O(M^{(L-1)/L} \sqrt N)$ queries.  Thus each of
the $t_2 = O((N/M)^{(L-1)/2})$ iterations of $W^{t_1} P$ uses $O(r +
t_1 \times M)=\tilde O(M^{(L-1)/L} \sqrt N + \sqrt M \times M)$
queries, where $M$ queries are needed for each walk step as in the
simple algorithm above.  Therefore, since $O(M^2)$ queries are
required for initialization, the total number of queries is
\be
  \tilde O(M^2 + (N/M)^{(L-1)/2} (M^{(L-1)/L} \sqrt N + M^{3/2}))
\,,
\label{eq:recursive}
\ee
and choosing $M=\nint{N^{L/(L+2)}}$ gives an overall query complexity
of $\tilde O(N^{(5L-2)/(2L+4)})$.

The numbers of queries used by these algorithms for $2 \le L \le 7$
are summarized in Table \ref{tab:cliques}.  Of course the algorithms
are not really specific to finding cliques, but could be used to find
any desired subgraph consisting of $L$ vertices.

In all the applications mentioned in this section so far (with $L \ge
3$), the problem has some structure that may allow us to learn
something about whether a given subset of fewer than $L$ items could
be part of an $L$-subset satisfying $\P$.  For example, in $L$-element
distinctness, if we find two inputs with different outputs, we know
these inputs cannot be part of an $L$-subset of inputs that all give
the same output.  Similarly, if we find two vertices in a graph that
are not connected by an edge, then we know that no triangle can
include these two vertices.  However, such structure is not used by
the algorithm; it can just as well solve a problem in which it is
impossible to obtain any information about whether fewer than $L$
inputs might satisfy $P$.  For example, consider the $L$-zero sum
problem with $D=\{1,\ldots,N\}$, $R=\{0,1\}^m$, and
$\P=\{((x_1,y_1),\ldots,(x_L,y_L)):~ y_1 \oplus \cdots \oplus y_L =
0^m\}$; or alternatively, suppose $R=\{0,1,\cdots,q-1\}$ and
$\P=\{((x_1,y_1),\ldots,(x_L,y_L)):~ y_1 + \cdots + y_L \equiv 0 \pmod
q\}$.  These particular $L$-subset finding problems seem to be
examples of the hardest cases, since there is no way to determine
anything about whether a subset of $L-1$ inputs might be part of an
$L$-subset satisfying $\P$.

\section{Open problems}
\label{sec:problems}

The subset finding algorithm we have described is based on a discrete
time quantum walk, but there is a more natural formulation of quantum
walk as a continuous time process that avoids the need to introduce a
coin register~\cite{FG98}.  For the present application the coin
register seems to be essential since it tells us what to query when we
take a step.  Nevertheless, we could consider a continuous time
quantum walk (without a coin) on the Johnson graph $J(N,M)$ whose
vertices are all subsets $A \subset D$ of size $|A|=M$, and in which
two vertices are connected if they differ in exactly one element.  If
we label the vertices $A$ of $J(N,M)$ with the corresponding black box
function values $f(A)$, and if we could simulate the continuous time
quantum walk on this labeled graph efficiently enough, then we could
have an algorithm that works in a similar way to the search algorithm
in~\cite{CG03}.  However, most of the techniques we currently know for
simulating such a walk are based on edge
coloring~\cite{Chi03,AT03,Lin87,Lin92} and do not work well when the
degree is large.  It is an open problem whether the walk on this graph
can be simulated efficiently enough to give a competitive algorithm
for the $L$-subset finding problem.

It would be interesting to know which of the algorithms we have
presented are optimal.  For the $L=1$ problem there is a well-known
lower bound of $\Omega(\sqrt N)$~\cite{BBBV97} that matches the
performance of the subset finding algorithm.  Of course for this
problem the simpler Grover algorithm also achieves this query
complexity~\cite{Gro97}.  For $L=2$ there in an $\Omega(N^{2/3})$
lower bound on element distinctness that follows by reduction from the
collision problem~\cite{Aar02,Shi01,Amb03a,Kut03}.  Again this matches
the performance of the subset finding algorithm~\cite{Amb03b}.
Therefore it is natural to conjecture an $\Omega(N^{L/(L+1)})$ lower
bound for the general $L$-subset finding problem.  To be concrete, we
propose
\begin{conjecture}
  The $L$-zero sum problem, i.e., the $L$-subset finding problem with
  domain $D=\{1,\ldots,N\}$, range $R=\{0,1\}^m$, and property
  $\P=\{((x_1,y_1),\ldots,(x_L,y_L)): y_1 \oplus \cdots \oplus y_L =
  0^m\}$, where $m$ is some function of $N$, requires
  $\Omega(N^{L/(L+1)})$ quantum queries.
\end{conjecture}
\noindent
While we believe this to be the case, the best lower bound we know is
$\Omega(N^{2/3})$, independent of $L$, for any $L \ge 2$.

Of course, it would also be interesting to prove lower bounds for
problems where something can be learned about subsets of size smaller
than $L$, such as $L$-element distinctness or finding an $L$-clique in
a graph.  For the former it is not so clear whether the algorithm
could be improved, whereas for the latter there is no particular
reason to expect that the known algorithms are optimal for all values
of $L$.

\medskip

{\em Note added.}  After this paper was posted to the quant-ph
archive, we learned through private communication that Magniez,
Santha, and Szegedy independently considered the problem of finding an
$L$-vertex subgraph in an $N$-vertex graph, and found an algorithm
using $\tilde O(N^{2(L-1)/L})$ queries.  This result is now described
in \cite{MSS03}.  In the context of the present paper, this can be
seen by choosing $M=\nint{N^{(L-1)/L}}$ in (\ref{eq:recursive}).  Note
that this result subsumes the simple algorithm (\ref{eq:simple}), and
also improves the $\tilde O(N^{(5L-2)/(2L+4)})$-query algorithm for
$L=5$.

\acknowledgments

We thank Scott Aaronson, Andris Ambainis, Wim van Dam, Edward Farhi,
Jeffrey Goldstone, Sam Gutmann, and Aram Harrow for helpful
discussions of the subset finding algorithm.  AMC also thanks Graeme
Ahokas and Richard Cleve for interesting discussions of the
implementation of continuous time quantum walks, and for providing
references~\cite{Lin87,Lin92}.

AMC received support from the Fannie and John Hertz Foundation, and
JME is supported by a National Science Foundation Graduate Research
Fellowship.  This work was also supported in part by the
Cambridge--MIT Institute, by the Department of Energy under
cooperative research agreement DE-FC02-94ER40818, and by the National
Security Agency and Advanced Research and Development Activity under
Army Research Office contract DAAD19-01-1-0656.


\bibliographystyle{apsrev_title}
\bibliography{subset}

\end{document}